\documentclass[fleqn,twoside,twocolumn,nofootinbib,showkeys]{revtex4} 
\usepackage[sec,nocpr]{ujp} 

\begin{document}
\title[Spectral Analysis and Invariant Measure]
{SPECTRAL ANALYSIS AND INVARIANT\\ MEASURE IN THE STUDY OF
 A NONLINEAR\\ DYNAMICS OF THE METABOLIC PROCESS IN CELLS}%
\author{V. Grytsay}
\affiliation{\bitp}
\address{\bitpaddr}

\udk{577.3} \pacs{05.45.-a, 05.45.Pq,\\[-3pt] 05.65.+b} \razd{\secx}

\autorcol{V.\hspace*{0.7mm}Grytsay}

\setcounter{page}{1}%

\begin{abstract}
The metabolic process in a cell is modeled with the use of the
Fourier transformation. The histograms of the invariant measures of
chaotic attractors are constructed. In particular, a scenario of
adaptation of the metabolic process under a change in the
dissipation of a kinetic membrane potential and the sequence of the
modes of self-organization and deterministic chaos are determined.
Respectively, the spectral mapping of the attractors of these modes
is considered. The structural-functional connections of the
metabolic process in a cell as an integral dissipative system are
analyzed.
\end{abstract}
 \maketitle

\section{Introduction}

In what follows, we will study the earlier constructed mathematical
model of the metabolic process in a cell
\cite{1,2,3,4,5,6,7,8,9,10,11,12,13,14,15,16,17,18,19}. The model is
based on the experimental data concerning bacteria
\textit{Arthrobacter globiformis} at a transformation of steroids
\cite{20}. This microorganism is referred to oxygen-breathing
bacteria arising 2.48~bln years ago. Due to the evolution of
metabolic processes in protobionts, the transition from the
oxygenless life of microorganisms to oxygen-breathing bacteria and
their subsequent evolution to eukaryotes happened. The relative
simplicity of the metabolic process running in given bacteria (the
absence of the Krebs cycle, e.g.) allows us to model the metabolic
process in a cell on the whole as an open dissipative structure, in
which two following basic systems necessary for the life are
self-organized: the system of a transformation of the substrate and
the respiratory chain. We will consider the specific biochemical
process of transformation of steroids by the given type of cells
\cite{2, 3}. This enables us to use the experimentally determined
parameters in the construction of a model and to make conclusion
about structural-functional connections under the self-organization
of the given biosystem. If some other substrate is used as a
nutrient medium, the mechanism of self-organization of a cell will
be analogous.

\section{Mathematical Model}

\begin{figure}%
\vskip1mm
\includegraphics[width=5cm]{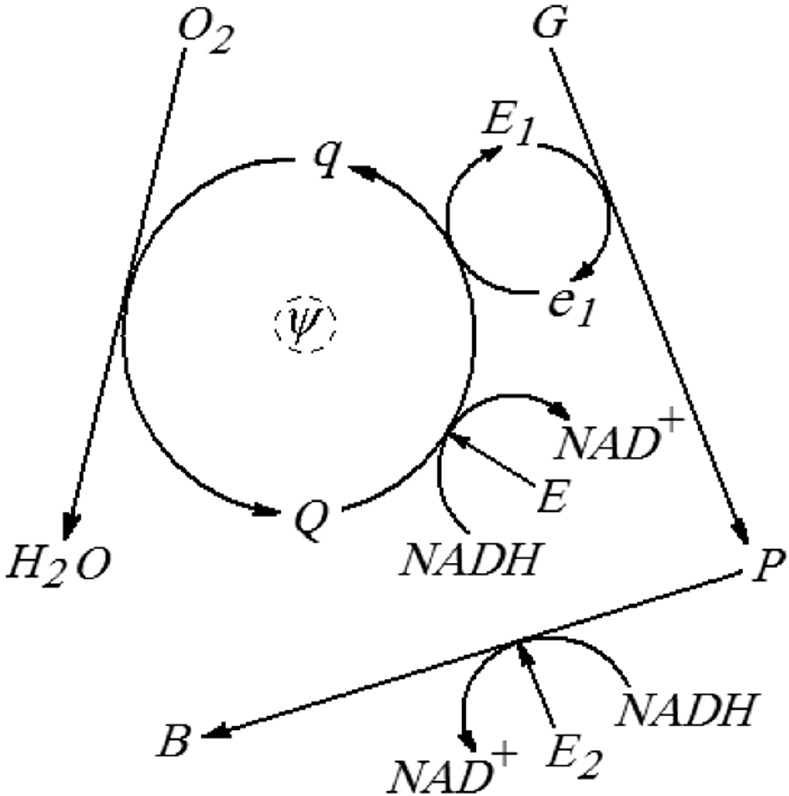}
\vskip-3mm\caption{General scheme of the metabolic process in a cell
}\label{fig:1}
\end{figure}

The mathematical model of the metabolic process in a cell is
constructed according to the general scheme of the process of
transformation of steroids by given cells (see Fig.~\ref{fig:1}) and
takes the form (\ref{eq1})--(\ref{eq10}):\vspace*{-1mm}
\begin{equation}\label{eq1}
\frac{dG}{dt}=\frac{G_0}{N_3+G+\gamma_2\psi}-l_1V(E_1)V(G)-\alpha_3G,
\end{equation}\vspace*{-7mm}
\begin{equation}\label{eq2}
\frac{dP}{dt}=l_1V(E_1)V(G)-l_2V(E_2)V(N)V(P) -\alpha_4P,
\end{equation}\vspace*{-7mm}
\begin{equation}\label{eq3}
\frac{dB}{dt}=l_2V(E_2)V(N)V(P)- k_1V(\psi)V(B) -\alpha_5B,
\end{equation}\vspace*{-7mm}
\[
\frac{dE_1}{dt}=E_{10}\frac{G^2}{\beta_1+G^2}\left(\!1-\frac{P+mN}{N_1+P+mN}\!\right)-
\]\vspace*{-9mm}
\begin{equation}\label{eq4}
-\,l_1V(E_1)V(G)+l_4V(e_1)V(Q) -a_1E_1,
\end{equation}
\begin{equation}\label{eq5}
\frac{de_1}{dt}=-l_4V(e_1)V(Q)+l_1V(E_1)V(G) -\alpha_1e_1,
\end{equation}\vspace*{-7mm}
\[
\frac{dQ}{dt}=6lV(2-Q)V(O_2)V^{(1)}(\psi)-l_6V(e_1)V(Q)\,-
\]\vspace*{-7mm}
\begin{equation}\label{eq6}
-\,l_7V(Q)V(N),
\end{equation}\vspace*{-7mm}
\begin{equation}\label{eq7}
\frac{dO_2}{dt}=\frac{O_{20}}{N_5+O_2}-lV(2-Q)V(O_2)V^{(1)}(\psi)-\alpha_7O_2,
\end{equation}\vspace*{-7mm}
\[
\frac{dE_2}{dt}=E_{20}\frac{P^2}{\beta_2+P^2}\frac{N}{\beta+N}\left(\!1-\frac{B}{N_2+B}\!\right)-
\]\vspace*{-7mm}
\begin{equation}\label{eq8}
-\,l_{10}V(E_2)V(N)V(P) -\alpha_2E_2,
\end{equation}\vspace*{-7mm}
\[
\frac{dN}{dt}=-l_2V(E_2)V(N)V(P)-l_7V(Q)V(N)\,+
\]\vspace*{-7mm}
\begin{equation}\label{eq9}
+\, k_2V(B)\frac{\psi}{K_{10}+\psi}+\frac{N_0}{N_4+N} -\alpha_6N,
\end{equation}\vspace*{-7mm}
\begin{equation}\label{eq10}
\frac{d\psi}{dt}=l_5V(E_1)V(G)+l_8V(N)V(Q) -\alpha\psi.
\end{equation}
where $V(X)=X/(1+X)$, $V^{(1)}(\psi )=1/(1+\psi ^2)$, $V(X)$ is a
function describing the adsorption of the enzyme in the region of a
local coupling, and $V^{(1)}(\psi )$ is a function characterizing
the influence of the kinetic membrane potential on the respiratory
chain.

In the modeling, it is convenient to use the following dimensionless
parameters: $l=l_1 =k_1 =0.2$; $l_2 =l_{10} =0.27$; $l_5 =0.6$; $l_4
=l_6 =0.5$; $l_7 =1.2$; $l_8 =2.4$; $k_2 =1.5$; $E_{10} =3$; $\beta
_1 =2$; $N_1 =0.03$; $m=2.5$; $\alpha =0.033$; $a_1 =0.007$; $\alpha
_1 =0.0068$; $E_{20} =1.2$; $\beta =0.01$; $\beta _2 =1$; $N_2
=0.03$; $\alpha _2 =0.02$; $G_0 =0.019$; $N_3 =2$; $\gamma _2 =0.2$;
$\alpha _5 =0.014$; $\alpha _3 =\alpha _4 =\alpha _6 =\alpha _7
=0.001$; $O_{20} =0.015$; $N_5 =0.1$; $N_0 =0.003$; $N_4 =1$;
$K_{10} =0.7$.

Equations (\ref{eq1})--(\ref{eq9}) present the changes in the
concentrations of: (\ref{eq1}) hydrocortisone ($G$); (\ref{eq2})
prednisolone ($P$); (\ref{eq3}) $20\beta $-oxyderivative of
prednisolone ($B$); (\ref{eq4}) oxidized form of
3-ketosteroid-$\Delta '$-dehydrogenase ($E_1 $); (\ref{eq5}) reduced
form of 3-ketosteroid-$\Delta '$-de\-hyd\-ro\-ge\-nase ($e_1 $);
(\ref{eq6}) oxidized form of the respiratory chain ($Q$);
(\ref{eq7}) oxygen ($O_2 $); (\ref{eq8}) 20$\beta
$-oxysteroid-dehydrogenase ($E_2 $); and (\ref{eq9}) ${\rm NAD}\cdot
H$ (reduced form of nicotinamide adenine dinucleotide) ($N)$.
Equation (\ref{eq10}) describes the change in a level of the kinetic
membrane potential ($\psi )$.

The reduction of parameters of the system to the dimensionless form
was performed in \cite{2,3}.

The presented mathematical model (\ref{eq1})--(\ref{eq10}) is
studied with the help of the application of the theory of nonlinear
differential equations \cite{21, 22} and the methods of simulation
of biochemical systems developed by the author and other researchers
in
\cite{23,24,25,26,27,28,29,30,31,32,33,34,35,36,37,38,39,40,41,42,43,44,45,46,47,48,49,50,51,52}.

This autonomous system of nonlinear differential equations was
solved by the Runge--Kutta--Merson method. The exactness of a
solution was set to be $10^{-12}$. To ensure the reliability of
calculations, namely, the passage of the system from the transient
initial phase to the asymptotic solution presented by an attractor,
the duration of calculations was taken to be $10^5$. In that time,
the trajectory ``sticks'' to the corresponding attractor.

\begin{figure*}%
\vskip1mm
\includegraphics[width=16.7cm]{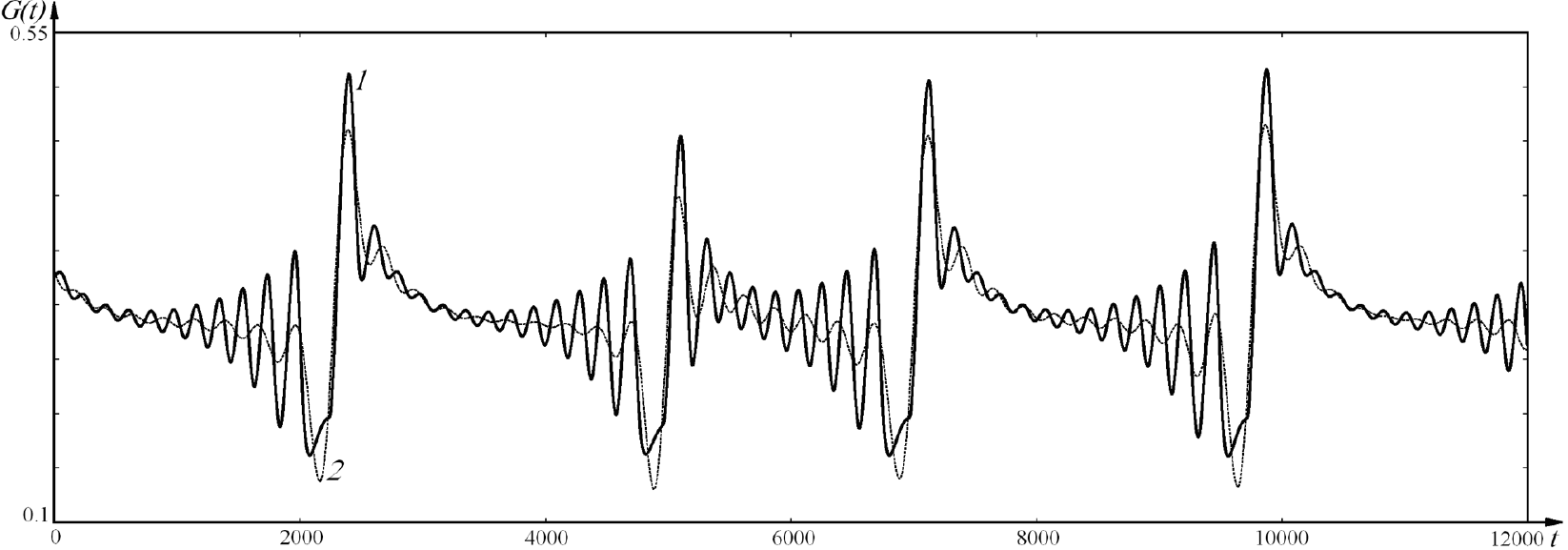}
\vskip-3mm\caption{Summary plot of the kinetics of the variable $G
$in the mode with the strange attractor $13^*2^x$ ($\alpha
=0.03217$) formed by the summation of harmonics: $n=1000$
(\textit{1}) and $n=200$ (\textit{2}) }\label{fig:2}\vspace*{-2mm}
\end{figure*}

The spectral analysis of the nonlinear dynamics of a metabolic
process was carried out by the expansion of the functions describing
the kinetics of the system in a trigonometric Fourier series in one
of the variables ($G$). Since we consider the autoperiodic
trajectories of regular attractors and strange attractors possessing
the fractality, the expansion was performed on the segment $[-l;
l]$. This means that the arbitrarily large period of any
autooscillatory mode of the system can be placed on the taken
segment $2l$ in length. The formula of such expansion in a Fourier
series reads \cite{53}:\vspace*{-3mm}
\[
a_0 +\sum\limits_{n=1}^\infty \left(\!(a_n \cos \frac{n\pi x}{l}
+b_n \sin \frac{n\pi x}{l}\!\right)\!,
\]
where\vspace*{-3mm}
\[
a_0 =\frac{1}{2l}\int\limits_{-l}^l {f(x)d(x)} , \quad a_n
=\frac{1}{l}\int\limits_{-l}^l {f(x)\cos \frac{n\pi x}{l}d(x)}
,\]\vspace*{-6mm}
\[
b_n =\frac{1}{l}\int\limits_{-l}^l {f(x)\sin \frac{n\pi x}{l}d(x)} ,
\quad n\in N.
\]
The spectrum of Lyapunov indices was calculated with the help of a
program written by the author in Fortran. Benettine's algorithm with
orthogonalization of vectors by the Gram--Schmidt method was used
\cite{21}.

As a quantitative measure of the fractality of strange attractors,
their Lyapunov dimension was calculated by the Kaplan--Yorke formula
\cite{54, 55}:
\[
D_{F_r } =m+\frac{\sum_{i=1}^m {\lambda _i } }{\left| {\lambda
_{m+1} } \right|}.
\]
\section{Results of Studies}

While studying the phase-parametric diagram of the system under
study, we have got the scenario of adaptation modes of the metabolic
process in a cell at a decrease in the dissipation of the kinetic
membrane potential $\alpha $ \cite{17}. The calculated plots of the
kinetics of modes manifest a stationary behavior and autoperiodic or
chaotic oscillations. They reflect the internal dynamics of the
metabolic processes in a cell. The possibility of the appearance of
autooscillatory modes in the given population of cells was later
confirmed experimentally \cite{56}.

In the present work in order to restrict its volume, we will
consider only separate modes, which does not affect the results.
Below, we will consider the scenario of appearance and destroying of
autooscillatory modes.

Stationary state $\mapsto 1^*2^0(\alpha =0.04131) \mapsto 2^*
2^0(\alpha =0.03753) \mapsto   3^* 2^0(\alpha =0.03563274) \mapsto
5^* 2^0(\alpha  =0.03463)  \mapsto   8^* 2^0(\alpha  =0.033) \mapsto
8^* 2^x(\alpha  =0.0328709)  \mapsto   11^* 2^0(\alpha =0.032516)
 \mapsto   11^* 2^x(\alpha  =0.03239) \mapsto  13^* 2^0(\alpha
 =0.03225) \mapsto  13^* 2^x(\alpha  =0.03217) \mapsto  7^* 2^x\mapsto
  1^* 2^{22}(\alpha  =0.032161)  \mapsto   1^* 2^0(\alpha  =0.0321148)
 \mapsto $ Stationary state.

In order to construct and compare the Fourier spectra on a single
scale, the value of $l$ was chosen with regard for the maximally
possible reasonable time of representation of the kinetics of the
most complicated mode, namely the strange attractor $13^*2^x$. In
this case, the number of harmonics of the expansion was taken:
$n=1000$. The kinetics of the given mode after the summation of all
harmonics is shown in Fig.~\ref{fig:2}, curve~\textit{1}. It
coincides completely with the initial plot (before the expansion) of
the kinetics of the given variable. For the value $n=200$, the
summary plot~2 of harmonics does not coincide with the initial one,
which means the smallness of the taken number of harmonics. In what
follows, the number of harmonics in expansions in a Fourier series
for any mode will be $n=1000$.

\begin{figure*}%
\vskip1mm
\includegraphics[width=16.7cm]{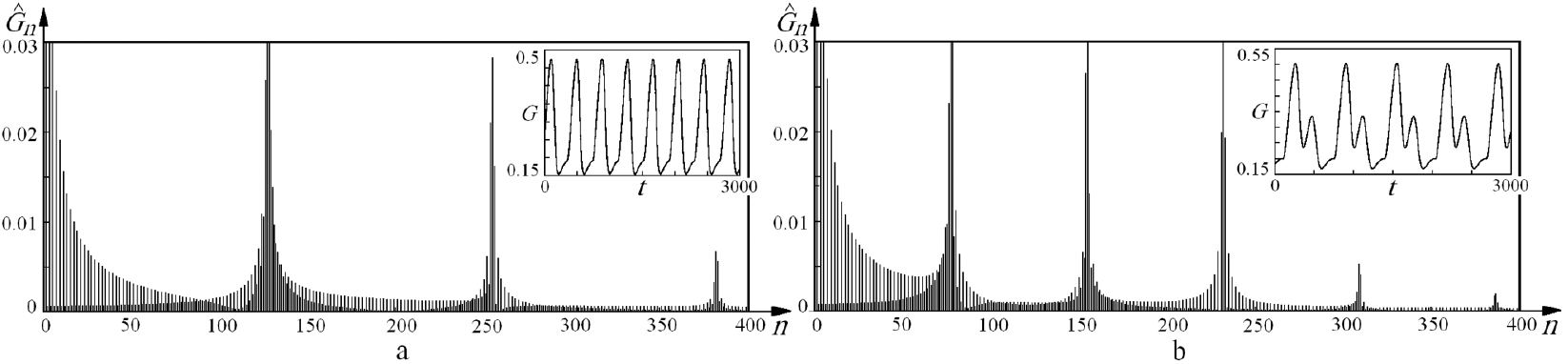}
\includegraphics[width=16.7cm]{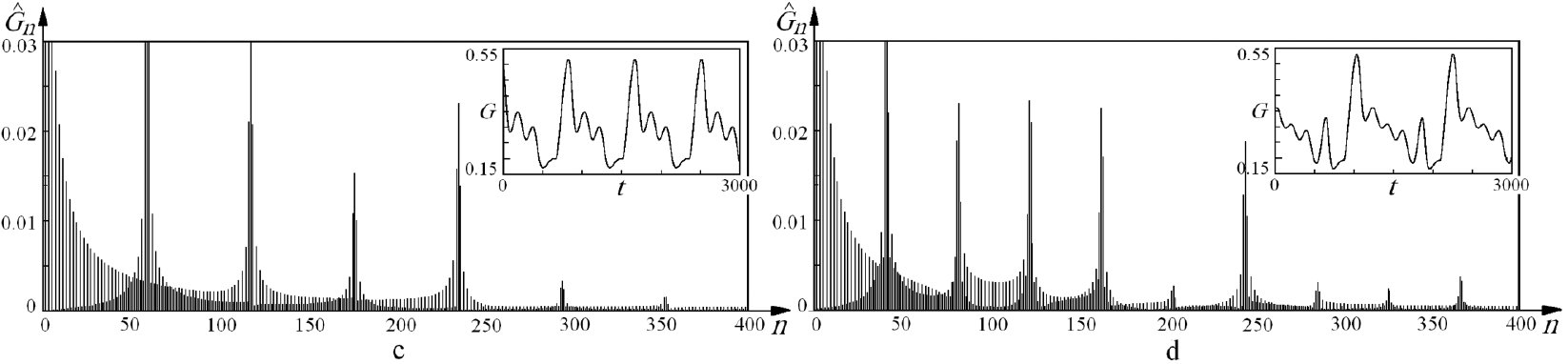}
\includegraphics[width=16.7cm]{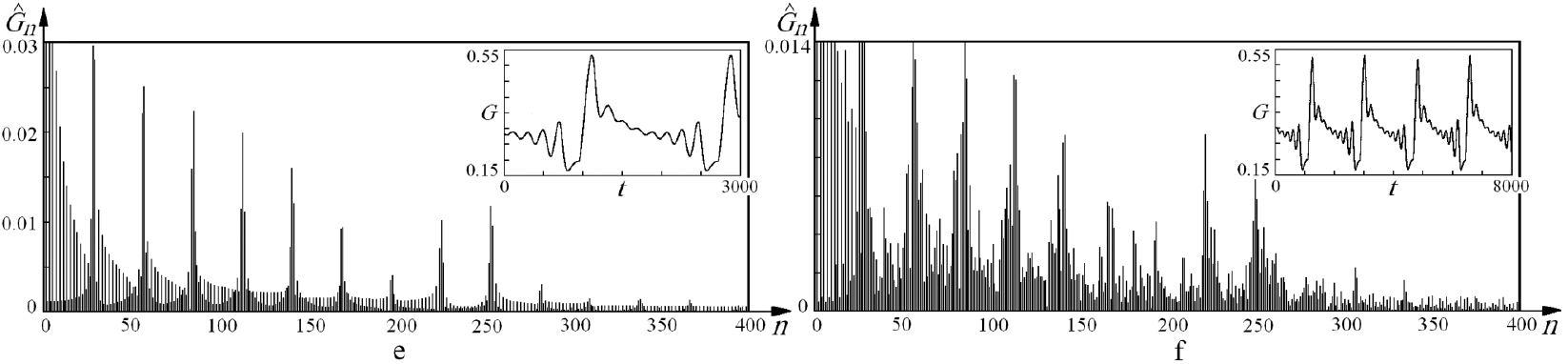}
\vskip-3mm\caption{Distribution of harmonics of the Fourier spectrum
in modes of the metabolic process: $a$ -- regular attractor
$1^*2^0(\alpha =0.04131$); $b$~-- regular attractor $2^*2^0 (\alpha
=0.03753$); $c$~-- regular attractor $3^*2^0 (\alpha =0.03563274$);
$d$~-- regular attractor $5^*2^0(\alpha =0.03463$); $e$~-- regular
attractor $8^*2^0(\alpha =0.033$); $f$~-- strange attractor $8^*2^x
(\alpha =0.0328709$)
  }\label{fig:3}\vspace*{-1mm}
\end{figure*}

\begin{figure*}%
\vskip1mm
\includegraphics[width=16.7cm]{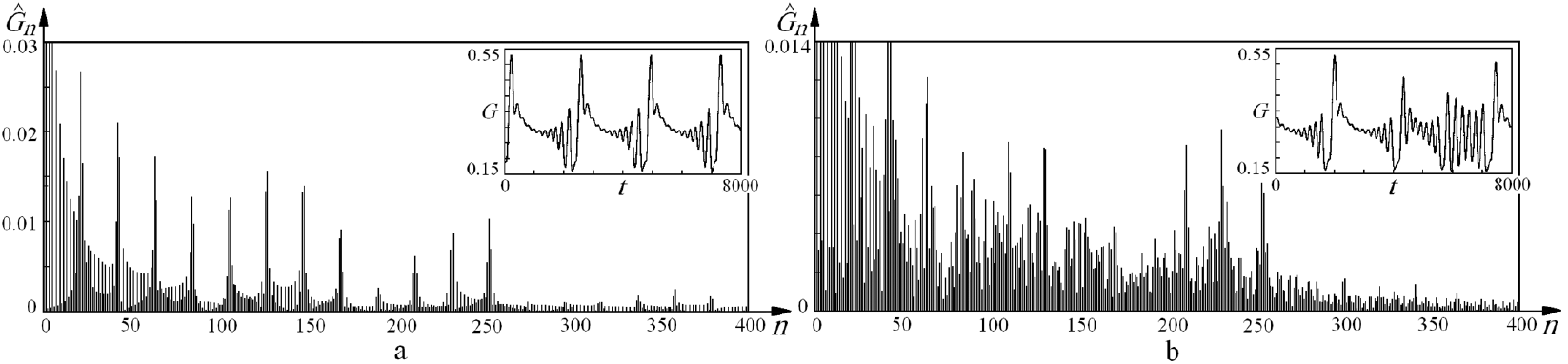}
\includegraphics[width=16.7cm]{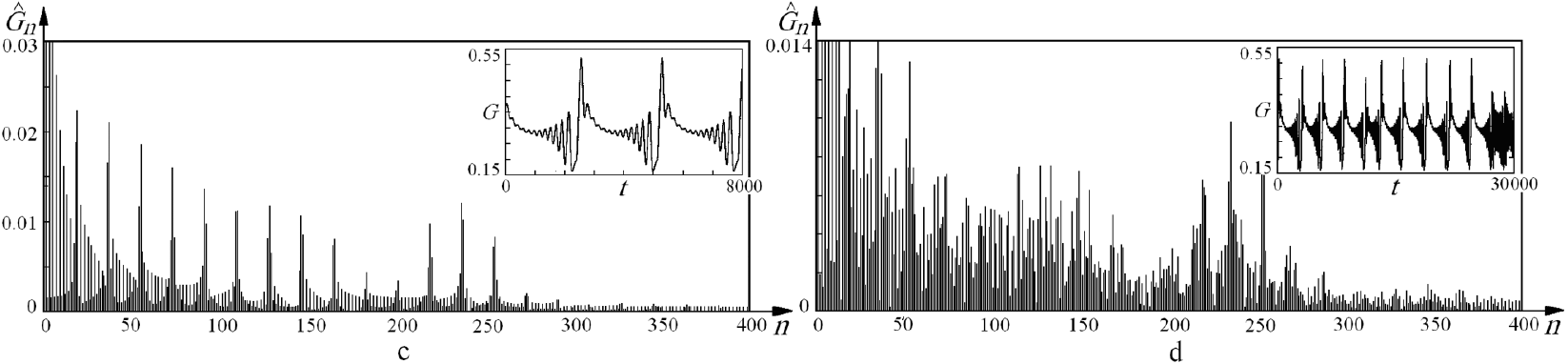}
\includegraphics[width=16.7cm]{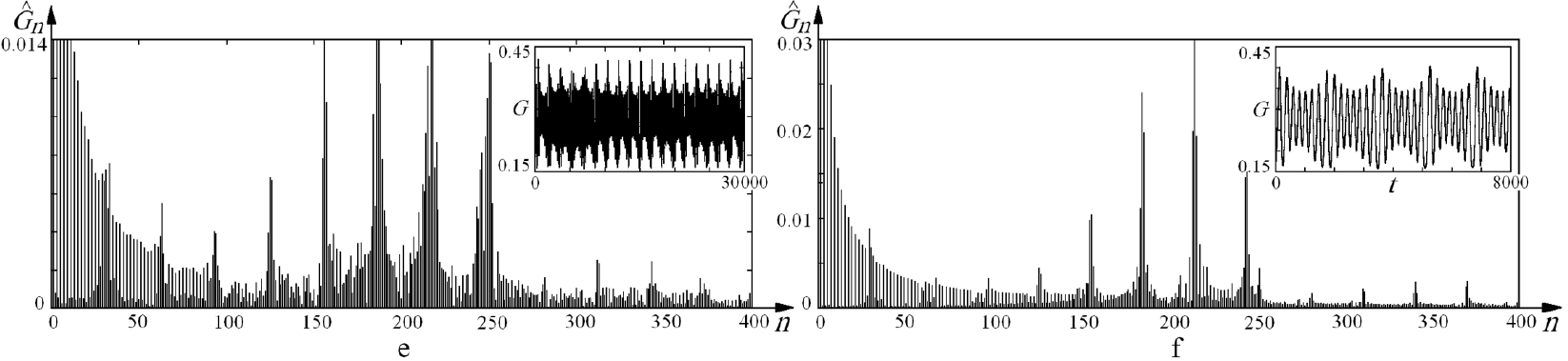}
\includegraphics[width=16.7cm]{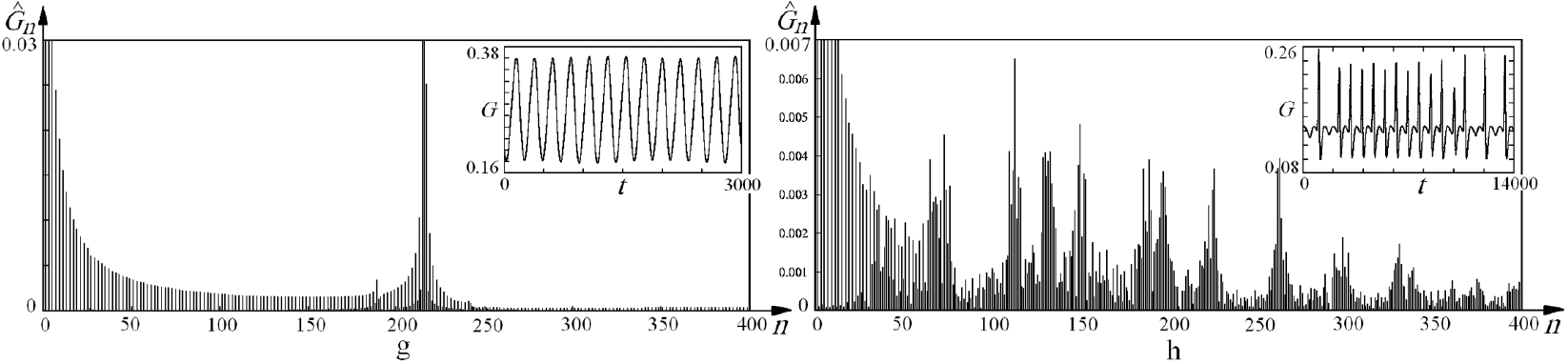}
\vskip-3mm\caption{Distribution of harmonics of the Fourier spectrum
in modes of the metabolic process: $a$~-- regular attractor $11^*2^0
(\alpha =0.032516$); $b$~-- strange attractor $11^*2^x(\alpha
=0.03239$; $c$~-- regular attractor $13^*2^0(\alpha =0.03225$);
$d$~-- strange attractor $13^*2^x (\alpha =0.03217$); $e$~-- strange
attractor $7^*2^x (\alpha =0.0321646)$; $f$~-- regular attractor
$1^*2^{22} (\alpha =0.032161$); $g$~-- regular attractor $1^*2^0
(\alpha =0.0321148$); $h$~-- strange attractor $3^*2^x (\alpha
=0.033$; $G_0 =0.009$; $O_{2_0 } =0.00209)$
}\label{fig:4}\vspace*{-2mm}
\end{figure*}

In Fig.~\ref{fig:3},~\textit{a--f} and \ref{fig:4},~\textit{a--g},
we show the spectra of expansion in a Fourier series of the variable
$G$ for some modes by the above-presented scenario for the variable
coefficient $\alpha $. In Fig.~\ref{fig:4},~\textit{h}, we give the
spectrum of some mode of a strange attractor for the input
parameters $G_0 =0.009$; $O_{2_0 } =0.00209$. In the plots, we
present the ratio of the amplitudes of harmonics $\hat{G}_n  $
bounded by a level of 0.03 for the sake of clearness for regular
attractors and by a level of 0.014 for strange attractors. Though
the calculation of each mode was made for $n=1000$, the plots
present only 400 harmonics. All harmonics for $n>400$ are
insignificant and omitted. In the right upper corner of each plot,
we give the kinetic curve of the given variable in this mode.

The plots given  correspond to the distribution of spectra of a
representation of the corresponding modes. As is seen, a decrease in
the dissipation of the kinetic membrane potential in the cyclic
metabolic process implies that some harmonics grow and become
maximal, whereas other ones decay to a minimum, on the contrary.
This complicates the nonlinear dynamics of the process. The decrease
in $\alpha $ from 0.04131 to 0.03753 causes a bifurcation and the
appearance of the 1-fold (Fig.~\ref{fig:3},~\textit{a}) and two-fold
(Fig.~\ref{fig:3},~\textit{b}) periodic modes of a regular
attractor. Respectively, this is revealed in the distribution of
harmonics.  The peaks of the basic harmonics of the expansion vary,
and one more peak has arisen. The further decrease in the
dissipation of the kinetic membrane potential leads to the
appearance of subsequent bifurcations. At $\alpha =0.03563274$, the
3-fold periodic mode (Fig.~\ref{fig:3},~\textit{c}) is formed, the
distribution of harmonics is changed, and one more peak appears. As
$\alpha $ decreases further, the 4-fold periodic mode arises
according to the scenario presented in [17]. Then the 5-fold cycle
is formed (Fig.~\ref{fig:3},~\textit{d}). Further as a result of
bifurcations, the 6-, 7-, and 8-fold periodic cycles appear
successively. The spectral distribution for the 8-fold cycle is
shown in Fig.~\ref{fig:3},~\textit{e}. We should like to indicate a
change in the spectral pattern at the given time moment. The further
decrease in the coefficient of dissipation of the kinetic membrane
potential does not lead to the appearance of a bifurcation and the
birth of a new cycle, but to the creation of the chaotic mode of the
strange attractor (Fig.~\ref{fig:3},~\textit{f}). The Fourier
distribution spectrum becomes more continuous. However, this
continuous spectrum includes clearly the harmonics of disappeared
limiting cycles. The transition between modes has arisen as a result
of the intermittence at the fracture of the laminar part of a
trajectory of the 8-fold limiting cycle by the turbulence. The
laminar part of the trajectory is formed at the expense of harmonics
of the regular attractor $8^*2^0$, whereas the turbulence is formed
due to the creation of new harmonics. The domain of attraction of
the limiting set of the regular attractor is eroded by these
harmonics. The further decrease in the coefficient $\alpha $ causes
the successive formation of the attractors $9^*2^0$, $9^*2^x$,
$10^*2^0$, $10^*2^x$, and $11^*2^0$.

The spectral analysis of the last regular attractor is shown in
Fig.~\ref{fig:4},~\textit{a}. It is seen how the self-organization
results in a change in the main harmonics of the expansion and in
their magnitudes and frequencies. As $\alpha $ decreases to 0.03239,
the self-organization of the given mode is violated, and the chaotic
mode of the strange attractor $11^*2^x$ is established
Fig.~\ref{fig:4},~\textit{b}. As in the previous case, the positions
of the maximal peaks of the basic harmonics of the previous mode
$11^*2^0$ are conserved. In this case, their magnitudes vary, which
testifies to the conservation of the attracting set of the given
regular attractor, whereas the appeared additional harmonics violate
it slightly. Due to the conservation of maximal harmonics, the
trajectory is kept in the domain of attraction of the attractor. The
further decrease in $\alpha $ causes the successive formation of the
attractors: $12^*2^0$, $12^*2^x$, and $13^*2^0 (\alpha =0.03225$)
(Fig.~\ref{fig:4},~\textit{c}). The change in the coefficient of
dissipation down to 0.03217 causes the destroying of this cycle and
the formation of the strange attractor $13^*2^x$
(Fig.~\ref{fig:4},~\textit{d}). It is the most complicated
attractor, for which we searched firstly for the necessary number of
harmonics reliably representing the nonlinear dynamics of the
process (Fig.~\ref{fig:2}). As above, the intermittence causes the
destroying of the stable 13-fold periodic cycle. The laminar
trajectories of peak harmonics of the attractor $13^*2^0$ are mixed
with the trajectories of new arisen harmonics
creating the \mbox{turbulence.}

\begin{figure*}%
\vskip1mm
\includegraphics[width=15cm]{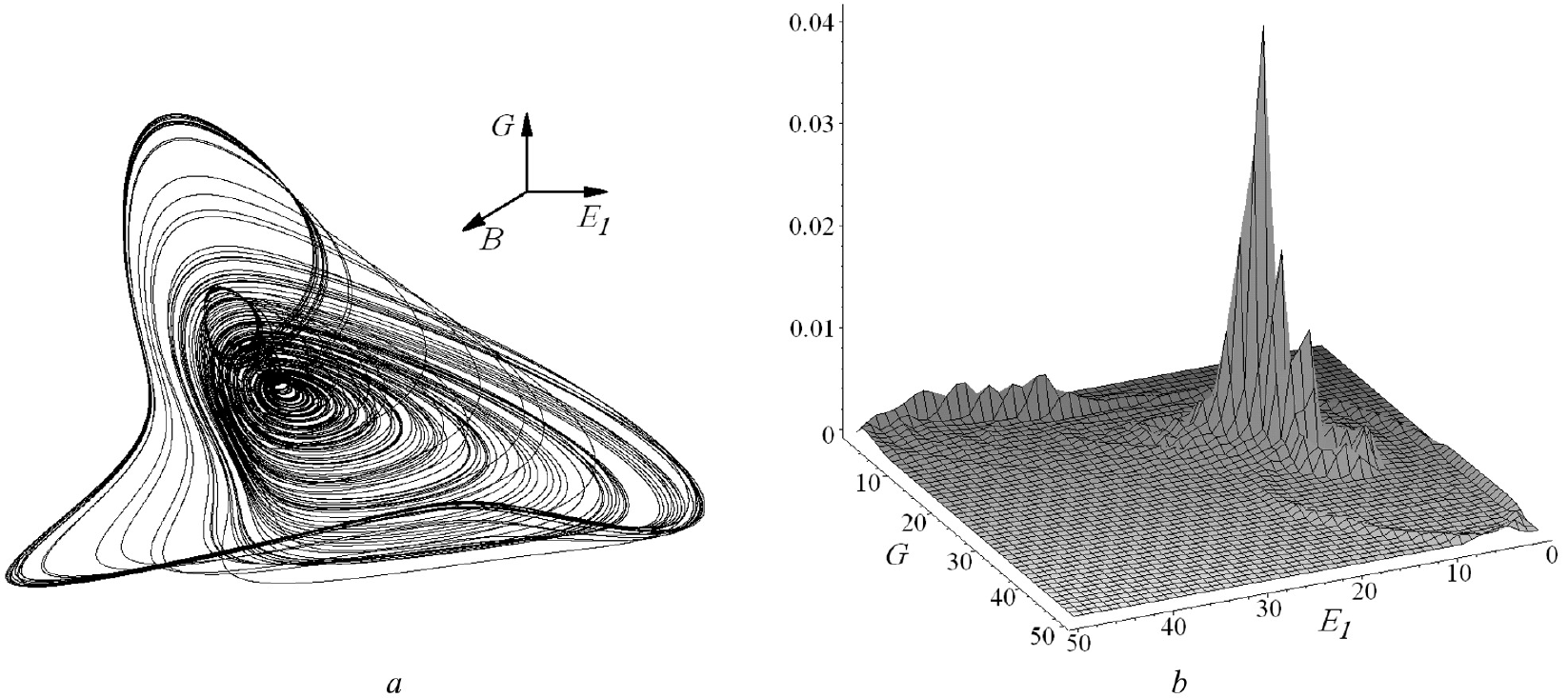}
\vskip-2mm\parbox{15cm}{\caption{Strange attractor $13^*2^x(\alpha
=0.03217$): \textit{a}~-- projection of its phase portrait in the
three-dimensional space $(E_1 ,G,B)$; \textit{b}~-- histogram of the
projection of the invariant measure on the plane $(G,E_1 )$, $t\in
(10^6$--$10^6+5\times 10^5)$\label{fig:5}}}
\end{figure*}

\begin{figure*}%
\vskip1mm
\includegraphics[width=15cm]{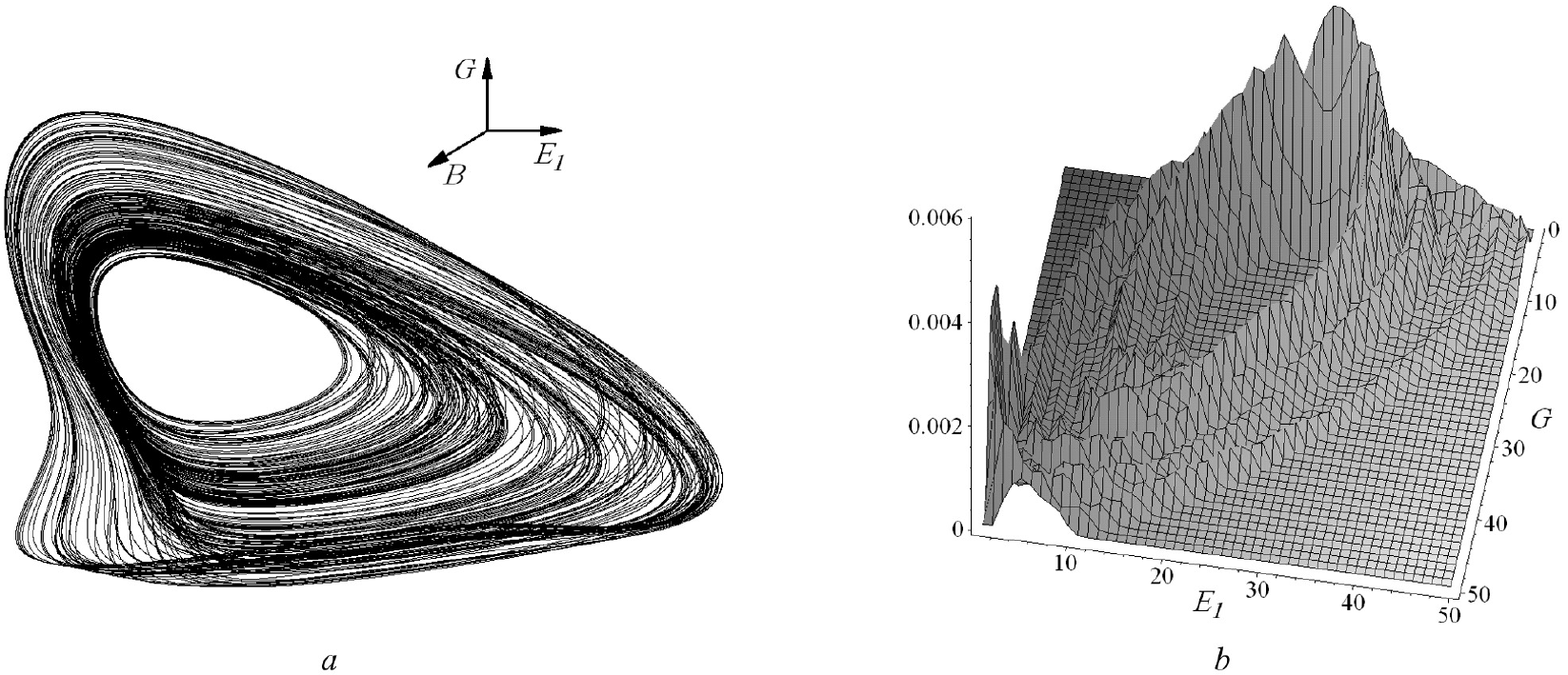}
\vskip-2mm\parbox{15cm}{\caption{Strange attractor $7^*2^x(\alpha
=0.0321646$): \textit{a}~-- projection of its phase portrait in the
three-dimensional space $(E_1 ,G,B)$; \textit{b}~-- histogram of the
projection of the invariant measure on the plane $(G,E_1 )$, $t\in
(10^6$--$10^6+5\times 10^5)$\label{fig:6}}}\vspace*{-1.5mm}
\end{figure*}

The subsequent decrease in $\alpha $ leads, strangely enough, to the
appearance of the unstable strange  attractor $7^*2^x$ ($\alpha
=0.0321646$) (Fig.~\ref{fig:4},~\textit{e}). As a result of the
self-organization, the basic bearing frequencies of harmonics vary,
and the different type of a 7-fold limiting cycle, which is the
attracting set for chaotic trajectories, is created. The further
decrease in $\alpha $ causes the subsequent self-organization and
the formation of a 22-fold autoperiodic cycle $1^*2^{22} (\alpha
=0.032161$) (Fig.~\ref{fig:4},~\textit{f}). Its expansion spectrum
contains a clearly expressed peak of the basic harmonic at a medium
frequency. The further decrease in the dissipation of the kinetic
membrane potential leads to the establishment of one dominant
frequency forming again a 1-fold periodic mode $1^*2^0(\alpha
=0.0321148$) (Fig.~\ref{fig:4},~\textit{g}). It passes to a
stationary mode, as $\alpha $ decreases. For the comparison of
modes, we show the distribution of harmonics of the Fourier spectrum
in Fig.~\ref{fig:4},~\textit{h} for the strange attractor $3^*2^x$
formed at $\alpha =0.033$; $G_0 =0.009$; $O_{2_0 } =0.00209$. The
spectrum of its basic bearing frequencies is essentially different
from that in the mode of the regular attractor $3^*2^0$
(Fig.~\ref{fig:3},~\textit{c}).

The study of the Fourier spectra of modes in
Figs.~\ref{fig:3},~\textit{a--f} and \ref{fig:4},~\textit{a--g}
testifies to a geometric similarity of the phase portraits of
attractors of the oscillatory dynamics of the system. The
redistribution of the amplitudes of harmonics is related to the
domains of attraction of that or other attractor. The bearing
frequencies characterize the laminar trajectories depending on the
multiplicity and the geometric type of an attractor. Their
significant change means a change in the domain of attraction of the
attractor and, respectively, the geometric type of the attractor
Fig.~\ref{fig:4},~\textit{h}.

In Fig.~\ref{fig:5},~\textit{a},~\textit{b}, we show a projection of
the phase portrait of the strange attractor $13^*2^x(\alpha
=0.03217)$ in the three-dimensional phase space $(E_1 ,G,B)$ and the
histogram of the projection of its invariant measure on the plane
$(G,E_1 )$.

This strange attractor is formed due to the funnel effect. As is
seen from Fig.~\ref{fig:5},~\textit{a}, there exists a domain in the
phase space, where the mixing of trajectories occurs. An arbitrarily
small deviation affects the evolution of the trajectory, and the
deterministic chaos is formed. Analogous funnels are formed also for
other strange attractors: $8^*2^x $, $9^*2^x $, $10^*2^x $, $11^*2^x
$ and $12^*2^x $.

In Fig.~\ref{fig:6},~\textit{a--b}, we present the strange attractor
$7^*2^x (\alpha =0.0321646$). In this case, as distinct from the
previous phase portrait, the funnel of mixing of trajectories is the
domain of instability increased to the size of the whole strange
attractor. It is a very unstable mode. As $\alpha $ increases
insignificantly, the funnel decreases, and the more stable strange
attractor $13^*2^x$ is established. But if $\alpha $
insignificantly, the self-organization results in the appearance of
a stabler regular attractor $1^*2^{22}$.

\begin{figure*}%
\vskip1mm
\includegraphics[width=15cm]{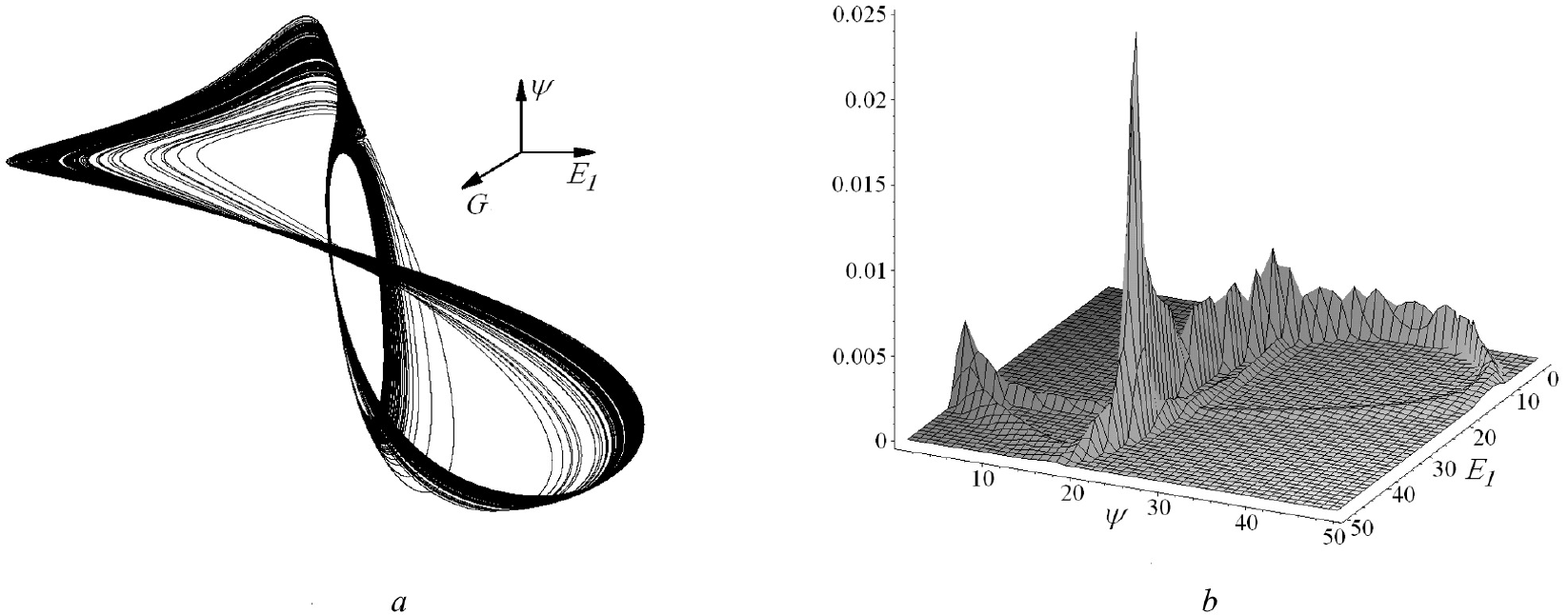}
\vskip-2mm\parbox{15cm}{\caption{Strange attractor $3^*2^x(\alpha
=0.033$; $G_0 =0.009$; $O_{2_0 } =0.00209)$: \textit{a}~--
projection of its phase portrait in the three-dimensional space
$(E_1 ,\psi ,G)$; \textit{b}~-- histogram of the projection of the
invariant measure on the plane $(E_1 ,\psi )$, $t\in
(10^6$--$10^6+2\times 10^6)$\label{fig:7}}}
\end{figure*}

As distinct from the above-shown attractors,
Fig.~\ref{fig:7},~\textit{a},~\textit{b} presents a strange
attractor that is formed not by a change in the coefficient $\alpha
$, but by a variation of the input parameters $G_0 $ and $O_{2_0 }
$. The obtained strange attractor in Fig.~\ref{fig:7},~\textit{a} is
created as a result of the intermittence of two unstable cycles
$3^*2^x$ and $2^*2^x$ (see the kinetics of $G$ in
Fig.~\ref{fig:4},~\textit{h}). As a result of the composition of two
unstable trajectories, the uncertainty in the evolution of the
system \mbox{arises.}

The above-considered phase portraits of the nonlinear dynamics are
typical of all studied modes of strange attractors of our system. In
Figs.~\ref{fig:5},~\textit{b},  \ref{fig:6},~\textit{b}, and
\ref{fig:7},~\textit{b}, we constructed the histograms of
projections of their invariant measures. To make it, the phase
spaces of the given strange attractors were partitioned into the
number of boxes, which is maximally possible for the computer
program, and the probabilities of the attendance of each box by the
trajectory were calculated. The numbers of points of the mapping for
the whole phase portrait were taken to be 500,000 in
Figs.~\ref{fig:5},~\textit{b} and \ref{fig:6},~\textit{b} and
2,000,000 in Fig.~\ref{fig:7},~\textit{b}. The numerical experiment
showed that a change in the numbers of points of the mapping has no
influence on the probability of the attendance of each box by the
trajectory. The time shift along a trajectory ($x(t)\to x(t+\tau ))$
has no influence as well, which means the invariance of the measures
of the given strange attractors.

\begin{table*}[!]
\vskip5mm \noindent\caption{Lyapunov's exponents of strange
attractors of the system}\vskip3mm\tabcolsep13.7pt

\noindent{\footnotesize\begin{tabular}{|c|c|c|c|c|c|c|c| }
 \hline \multicolumn{1}{|c}
{\rule{0pt}{5mm}$\alpha $} & \multicolumn{1}{|c}{Attractor}&
\multicolumn{1}{|c}{$\lambda _1 $}& \multicolumn{1}{|c}{$\lambda _2
$}& \multicolumn{1}{|c}{$\lambda _3 $}& \multicolumn{1}{|c}{$\lambda
_{10} $}& \multicolumn{1}{|c}{$\Lambda $}&
\multicolumn{1}{|c|}{$D_{F_r} $}\\[1.5mm]%
\hline%
\rule{0pt}{4mm}0.032872~\,& ~\,$8^*2^x$ & 0.00013& 0.00000&
--0.00506& --0.52172& --0.93352&
--2.02569 \\
 0.03268~\,~\,& ~\,$9^*2^x$& 0.00049& 0.00000& --0.00500& --0.51658&
--0.92495&
--2.0980~\, \\
 0.03254~\,~\,& $10^*2^x$& 0.00026& 0.00000& --0.00459& --0.50965&
--0.91713&
--2.05664 \\
 0.03242~\,~\,& $11^*2^x$& 0.00035& 0.00000& --0.00449& --0.50683&
--0.91260&
--2.07795 \\
 0.032278~\, & $12^*2^x$& 0.00060& 0.00000& --0.00402& --0.50460&
--0.90649&
--2.14925 \\
 0.03217~\,~\,& $13^*2^x$& 0.00083& 0.00000& --0.00361& --0.50326&
--0.90210&
--2.23075 \\
 0.0321646& ~\,$7^*2^x$& 0.00031& 0.00000& --0.00111& --0.51650&
--0.90393&
--2.27928 \\[2mm]
\hline
\end{tabular}}\vspace*{-2.5mm}
\end{table*}

The peak of the invariant measure in Fig.~\ref{fig:5},~\textit{b}
characterizes visually the attracting set of the given attractor in
the funnel, and the ability to mix its trajectories in this
compressed domain of the phase space. Figure~\ref{fig:6},~\textit{b}
contains no such domain. The mixing occurs over the whole domain of
the phase \mbox{space.}\looseness=1%

For the above-considered scenario of formation and destroying of
strange attractors, we calculated the complete spectra of Lyapunov's
exponents and the Lyapunov dimensions of their fractality (Table)
(indices $\lambda _4 -\lambda _9 $ are omitted, since they do not
influence the meaning of the presented results).

According to the data presented in Table, the KS-entropy
(Kolmogorov--Sinai entropy) \cite{57} of the strange attractor
$13^*2^x$ is maximal and equals 0.00083. For the attractor $7^*2^x$,
it is equal to 0.00031. This means that the exponential divergence
of phase trajectories along the vector of perturbations
corresponding to $\lambda _1 $ for the strange attractor $13^*2^x$
is larger than for the attractor $7^*2^x$. For the rest vectors
corresponding to negative values of $\lambda _3 -\lambda _{10} $,
the phase trajectories approach exponentially their attractor.
Respectively, the nonpredictability and chaoticity of the
deterministic chaos for the first strange attractor are higher than
for the second one. The larger the value of KS-entropy, the more
complicated is the structure of a chaos. At the same time, the ratio
of the Lyapunov dimensions of the fractality of these modes is
opposite. We have $D_{F_r} =-2.23075$ for the strange attractor
$13^*2^x$ and $D_{F_r} =-2.27928$ for $7^*2^x$. This can be
explained by that the Lyapunov dimensions for these modes are
defined not only by the values of $\lambda _1 $ and $\lambda _2 $,
but also by $\left| {\lambda _3 } \right|$. This characterizes a
deformation of an element of the phase volume along the
corresponding vectors of perturbations. The larger $\lambda _1 $ and
the less $\left| {\lambda _3 } \right|$, the larger is the
deformation of its \mbox{volume.}

The studies showed that the domain of instability of autooscillatory
modes is located between two stationary modes of the metabolic
process. It appears due to the breaking of a synchronization between
the system of consumption of a substrate and the respiratory chain.
This breaking occurs due to the expenditure of the proton membrane
potential of a cell for other metabolic processes. A decrease in the
potential causes the desynchronization of the processes of
consumption of a substrate and the processes of transport and
accumulation of a charge on the outer side of the membrane. The
blocking of the respiratory chain by an increased level of the
potential is decreased, the desynchronization arises, and
autoperiodic or chaotic oscillations in the metabolic process
appear.

The executed expansions in a Fourier series of the determined
autooscillatory modes allow us to substantiate this method for the
seeking and the identification of autoperiodic and chaotic modes in
the metabolic process in a cell. The calculated histograms of the
invariant measures of strange attractors give a more complete visual
representation of the domains, where the trajectories are mixed, as
compared with the phase portraits.\vspace*{-1mm}

\section{Conclusions}

In the frame of the mathematical model the autooscillatory modes of
the metabolic process in a cell are studied with the help of the
expansion of the kinetics of the process in a Fourier series and the
construction of histograms of the invariant measures of chaotic
attractors of the model. We determined the necessary number of
harmonics, which represent uniquely the most complicated mode of a
strange attractor. The dependence of the type of an attractor on the
distributions and the amplitudes of harmonics in the Fourier
spectrum is investigated The harmonics forming the laminar and
turbulent parts of a trajectory of the attractor are identified. The
histograms of projections of the invariant measures of the main
types of strange attractors of the system are constructed. Their
dependence on the phase portrait is determined. The complete spectra
of Lyapunov's exponents, KS-entropy, and Lyapunov dimensions of the
fractality for the strange attractors under study are calculated.
The mechanisms and the causes for the appearance of autoperiodic and
chaotic oscillations in the metabolic process in a cell are
found.

\vskip2mm\textit{The work is supported by the project
No.\,0113U001093 of the National Academy of Sciences of Ukraine.}


\rezume{%
В.Й.\,Грицай}{СПЕКТРАЛЬНИЙ АНАЛIЗ ТА IНВАРIАНТНА\\ МIРА ПРИ
ДОСЛIДЖЕННI НЕЛIНIЙНОЇ ДИНАМIКИ\\ МЕТАБОЛIЧНОГО ПРОЦЕСУ В КЛIТИНІ}
{Проведено моделювання метаболiчного процесу в клітині з
використанням перетворення Фур'є та побудови гiстограм iнварiантних
мiр хаотичних атракторiв. Зокрема, знайдено сценарiй адаптацiї
метаболiчного процесу при змiнi дисипацiї кiнетичного мембранного
потенцiалу, послiдовнiсть режимiв самоорганiзацiї та детермiнованого
хаосу і, відповідно, розглянуто спектральне вiдображення з
атракторами цих режимiв. Проаналізовано структурно-функцiональнi
зв'язки метаболiчного процесу в клiтині як єдиної дисипативної
системи.}


\begin{thebibliography}{99}
\bibitem{1} V.P. Gachok, V.I. Grytsay.
Kinetic model of macroporous granule with the regulation of
biochemical processes. \textit{Dokl. Akad. Nauk SSSR} \textbf{282},
No.~1, 51 (1985).\vspace*{0.5mm}
\bibitem{2} V.P. Gachok, V.I. Grytsay, A.Yu. Arinbasarova, A.G.~Me\-dentsev, K.A. Koshcheyenko, V.K. Akimenko.
Kinetic model of hydrocortisone 1-en-dehydrogenation by
Arth\-ro\-bac\-ter globiformis. \textit{Biotechn. Bioengin.}
\textbf{33}, 661 (1989) [DOI:
10.1002/bit.260330602].\vspace*{0.55mm}
\bibitem{3} V.P. Gachok, V.I. Grytsay, A.Yu. Arinbasarova, A.G.~Me\-dentsev, K.A. Koshcheyenko, V.K. Akimenko.
Kinetic model for the regulation of redox reaction in steroid
transformation by Arthrobacter globiformis cells. \textit{Biotechn.
Bioengin.} \textbf{33}, 668 (1989) [DOI:
10.1002/bit.260330603].\vspace*{0.55mm}
\bibitem{4} V.I. Grytsay. Self-organization  in  the  macroporous
structure  of  the  gel  with immobilized  cells.  Kinetic  model of
bioselective  membrane  of biosensor.  \textit{Dopov. Nats. Akad.
Nauk Ukr.} No.~2, 175 (2000).\vspace*{0.55mm}
\bibitem{5} V.I. Grytsay. Self-organization  in  a  reaction-diffusion  porous  media.
\textit{Dopov. Nats. Akad. Nauk Ukr.} No.~3, 201
(2000).\vspace*{0.55mm}
 \bibitem{6} V.I. Grytsay.
Ordered  structure  in  a  mathematical model  biosensor.
\textit{Dopov. Nats. Akad. Nauk Ukr.} No.~11, 112
(2000).\vspace*{0.55mm}
\bibitem{7} V.I. Grytsay.
Self-organization  of  biochemical  process  of  immobilized  cells
bioselective of membrane biosensor.  \textit{Ukr. J. Phys.}
\textbf{46}, 124 (2001).\vspace*{0.55mm}
\bibitem{8} V.V. Andreev, V.I. Grytsay.
Modeling of inactive zones in porous granules of a catalyst and in a
biosensor.  \textit{Matem. Modelir.} \textbf{17}, No.~2, 57
(2005).\vspace*{0.55mm}
\bibitem{9} V.V. Andreev, V.I. Grytsay.
Influence of heterogeneity of diffusion-reaction process for the
formation of structures in the porous medium.
 \textit{Matem. Modelir.} \textbf{17}, No.~6, 3 (2005).\vspace*{0.55mm}
\bibitem{10} V.I. Grytsay, V.V. Andreev.
The  role  of  diffusion  in  the  active  structures formation in
porous reaction-diffusion media.  \textit{Matem. Modelir.}
\textbf{18}, No.~12, 88 (2006).\vspace*{0.55mm}
\bibitem{11} V.I. Grytsay.
Unsteady  conditions  in  porous  reaction-diffusion. \textit{Roman.
J. Biophys.} \textbf{17,} No.~1, 55 (2007).\vspace*{0.55mm}
\bibitem{12} V.I. Grytsay.
The  uncertainty  in  the  evolution  structure  of
reaction-diffusion medium bioreactor.  \textit{Biofiz. Visn.}
\textbf{19}, No.~2, 92 (2007).\vspace*{0.55mm}
\bibitem{13} V.I. Grytsay.
Morphogenetic field forming and stability of bioreactor
immobilization cells. \textit{Biofiz. Visn.} \textbf{20}, No.~1, 48
(2008).\vspace*{0.55mm}
\bibitem{14} V.I. Grytsay.
Prediction structural instability and type attractor of biochemical
process. \textit{Biofiz. Visn.} \textbf{23}, No.~2, 77
(2009).\vspace*{0.55mm}
\bibitem{15} V.I. Grytsay.
Structural instability of a biochemi cal process.  \textit{Ukr. J.
Phys.} \textbf{55}, No.~2, 599 (2010).\vspace*{0.55mm}
\bibitem{16} V.I. Grytsay, I.V. Musatenko.
Self-oscillatory
  dynamics  of  the  metabolic
process in a cell.  \textit{Ukr. Biochem. J.} \textbf{85}, No.~2, 93
(2013) [DOI: 10.15407/ubj85.02.093].\vspace*{0.55mm}
\bibitem{17} V.I. Grytsay, I.V. Musatenko.
  The  structure  of  a  chaos  of  strange  attractors
within a mathematical model of the metabolism of a cell.
\textit{Ukr. J. Phys.} \textbf{58}, No.~7, 677 (2013) [DOI:
10.15407/ujpe58.07.0677].\vspace*{0.55mm}
\bibitem{18} V. Grytsay, I. Musatenko.
A  mathematical  model of the  metabolism of  a cell.
Self-organisation and chaos. \textit{Chaotic Modeling and Simulation
(CMSIM)} No.~4, 539 (2013).
\bibitem{19} V.I. Grytsay, I.V. Musatenko.
Self-organization and chaos in the metabolism of a cell.
\textit{Biopolym. Cell} \textbf{30} No. 5, 403 (2014) [DOI:
10.7124/bc.0008B9].\vspace*{0.5mm}
\bibitem{20} A.A. Akhrem, Yu.A. Titov. \textit{Steroids and Microorganisms} (Nauka,  1970) (in Russian).\vspace*{0.5mm}
\bibitem{21} S.P. Kuznetsov. \textit{Dynamical Chaos} (Fizmatlit,  2001) (in Russian).\vspace*{0.5mm}

\bibitem{22} V.S. Anishchenko. \textit{Complex Oscillations in Simple Systems} (Nauka,
 1990) (in Russian).\vspace*{0.5mm}

\bibitem{23} Yu.M. Romanovskii, N.V. Stepanova, D.S. Chernavskii.
\textit{Mathematical Biophysics} (Nauka,  1984) (in
Rissian).\vspace*{0.5mm}

\bibitem{24} G.G. Malinetskii, A.B. Potapov. \textit{Modern Problems of
Nonlinear Dynamics} (Editorial URSS,  2002) (in
Russian).\vspace*{0.5mm}

\bibitem{25} E.E. Selkov. Self-Oscillations in Glycolysis. \textit{Europ. J. Biochem.}
\textbf{4}, 79 (1968) [DOI:
10.1111/j.1432-1033.1968.tb00175.x].\vspace*{0.5mm}

\bibitem{26} M. Holodniok, A. Klic, M. Kubicek, M. Marek. \textit{Methods of Analysis
of Nonlinear Dynamical Models} (Academia,  1986) (in
Czech).\vspace*{0.5mm}

\bibitem{27} G.Yu. Riznichenko. \textit{Mathematical Models in Biophysics and Ecology}
(Inst. of Computer. Studies,  2003) (in Russian).\vspace*{0.5mm}

\bibitem{28} V.S. Podgorskij. \textit{Physiology and Metabolism of
Methanol-Assimilating Yeast} (Naukova Dumka,  1982) (in
Russian).\vspace*{0.5mm}

\bibitem{29} V. Anishchenko, V. Astakhov, A. Neiman, T. Vadicasova, L.
Schimansky-Geir. \textit{Nonlinear Dynamics of Chaotic and
Stochastic System. Tutorial and Modern Developments} (Springer,
 2007) [ISBN: 978-3-540-38168-6].\vspace*{0.5mm}

\bibitem{30}  \textit{Chaos in Chemical and Biochemical System}. Ed. by R.~Field, L.~Gyorgyi (World Scientific Press,  1993).\vspace*{0.5mm}

\bibitem{31} V.A. Kordium, D.M. Irodov, O.O. Maslova, T.A. Ruban, E.M.
Sukhorada, V.I. Andrienko, N.S. Shuvalova, L.I. Likhachova, S.P.
Shpilova. Fundamental biology reached a plateau~-- development of
ideas. \textit{Biopolimers \& Cell} \textbf{27} (6), 480
(2011).\vspace*{0.5mm}

\bibitem{32} V.I. Grytsay. The conditions of the self-organization in the
multienzyme prostacyclin-thromboxane system. \textit{Visn. Kyiv.
Univ.} No.~3, 372 (2002).\vspace*{0.5mm}

\bibitem{33} V.I. Grytsay, V.P. Gachok. The modes of self-organization в
prostacyclin-thromboxane system.  \textit{Visn. Kyiv. Univ.} No.~4,
365 (2002).\vspace*{0.5mm}

\bibitem{34} V.I. Grytsay, V.P. Gachok. Ordered structures in the
mathematical system of prostacyclin and thromboxane model.
\textit{Visn. Kyiv. Univ., Ser. Fiz.-Mat. Nauk.} No.~1, 338
(2003).\vspace*{0.5mm}

\bibitem{35} V.I. Grytsay. Modeling of processes in the multienzyme
prostacyclin and thromboxane system.  \textit{Visn. Kyiv. Univ.}
No.~4, 379 (2003).\vspace*{0.5mm}

\bibitem{36} V.P. Gachok, Kinetics of Biochemical Processes. (Naukova Dumka,
Kiev, 1988) (in Russian).\vspace*{0.5mm}

\bibitem{37} V.P. Gachok. \textit{Strange Attractors in Biosystems} (Naukova Dumka,
 1989) (in Russian).\vspace*{0.5mm}

\bibitem{38} S.D. Varfolomeev, A.T. Mevkh, V.P. Gachok. Kinetic model of the
multienzyme system of blood prostanoid synthesis. 1. Mechanism of
stabilization of the levels of thromboxane and prostacyclin.
\textit{Molek. Biol.} \textbf{20}, No.~4, 957 (1986).\vspace*{0.6mm}

\bibitem{39} S.D. Varfolomeev, V.P. Gachok, A.T. Mevkh. Kinetic behavior of
the multienzyme system of blood prostanoid synthesis.
\textit{BioSystems} \textbf{19}, 45 (1986).\vspace*{0.6mm}

\bibitem{40} V.I. Grytsay, I.V. Musatenko. Self-organization and fractality
in metabolic processes of the Krebs cycle. \textit{Ukr. Biokhim.
Zh.} \textbf{85}, No.~5, 191 (2013).\vspace*{0.6mm}

\bibitem{41} V. Grytsay, I. Musatenko. Nonlinear self-organization dynamics
of a metabolic process of the Krebs cycle. \textit{Chaotic Modeling
and Simulation (CMSIM)} \textbf{3}, 207 (2014).\vspace*{0.6mm}

\bibitem{42} V. Grytsay. Lyapunov indices and the poincare mapping in a study
of the stability of the krebs cycle. \textit{Ukr. J. Phys.}
\textbf{60}, No.~6, 561 (2015) [DOI:
10.15407/ujpe60.06.0561].\vspace*{0.6mm}

\bibitem{43} V.I. Grytsay. Self-organization and fractality in the metabolic
process of glycolysis. \textit{Ukr. J. Phys.} \textbf{60}, No.~12,
1243 (2015) [DOI: 10.15407/ujpe60.12.1251].\vspace*{0.6mm}

\bibitem{44} V.I.Grytsay. Self-organization and chaos in the metabolism of
hemostasis in a blood vessel. \textit{Ukr. J. Phys.} \textbf{61},
No.~7, 648 (2016) [DOI: 10.15407/ujpe61.07.0648].\vspace*{0.6mm}

\bibitem{45} V.I. Grytsay. A mathematical model of the metabolic process of
atherosclerosis. Ukr. Biochem. J. 88, No. 4, 75 (2016) [DOI:
10.15407/ubj88.04.075].\vspace*{0.6mm}

\bibitem{46} V. Grytsay. Self-organization and fractality created by
gluconeogenesis in the metabolic process.\textit{ Chaotic Modeling
and Simulation (CMSIM)} \textbf{2}, 113 (2016).\vspace*{0.6mm}

\bibitem{47} A. Golub, O. Matyshevska, S. Prylutska, V. Sysoyev, L.~Ped, V.
Kudrenko, E. Radchenko, Yu. Prylutskyy, P.~Scharff, T. Braun.
Fullerenes immobilized at silica surface: topology, structure and
bioactivity. \textit{J. Mol. Liq.} \textbf{105}, No.~2--3, 141
(2003) [DOI: 10.1016/S0167-7322(03)\mbox{00044-8}].\vspace*{0.6mm}

\bibitem{48}  Yu.I. Prylutskyy, V.M. Yashchuk, K.M. Kushnir, A.A.~Go\-lub, V.A.
Kudrenko, S.V. Prylutska, I.I. Grynyuk, E.V.~Bu\-za\-neva, P.
Scharff, T. Braun, O.P. Matyshevska. Biophysical studies of
fullerene-based composite for bio-nanotechnology. \textit{Mater.
Sci. Engineer. C} \textbf{23}, No.~1--2, 109 (2003) [DOI:
10.1016/S0928-4931(02)00244-8].\vspace*{0.6mm}

\bibitem{49}  A.D. Suprun, Yu.I. Prylutskyy, A.M. Shut, M.S. Miroshnichenko.
Towards a dynamical model of skeletal muscle.\textit{ Ukr. J. Phys.}
\textbf{48}, No.~7, 704 (2003).\vspace*{0.6mm}

\bibitem{50} Yu.I. Prylutskyy, A.M. Shut, M.S. Miroshnychenko, A.D.~Sup\-run.
Thermodynamic and mechanical properties of skeletal muscle
contraction. \textit{Int. J. Thermophys.} \textbf{26}, No. 3, 827
(2005) [DOI: 10.1007/s10765-005-5580-8].\vspace*{0.6mm}

\bibitem{51} A.D. Suprun, A.M. Shut, Yu.I. Prylutskyy. Simulation of the Hill
equation for fibre skeletal muscle contraction. \textit{Ukr. J.
Phys.} \textbf{52}, No.~10, 997 (2007).\vspace*{0.6mm}

\bibitem{52} М. Zabolotnyy, Yu. Barabash, Yu. Sklyarov, Yu. Prylutskyy. The
model of photoinduced changes in the pigment-protein complex of
reaction center. \textit{Ukr. Bioorg. Acta} No.~1, 27
(2010).
\bibitem{53} N.S. Piskunov. \textit{Differential and Integral Calculi} (Nauka,  1978) (in Russian).\vspace*{0.5mm}
\bibitem{54} J.L. Kaplan, J.A. Yorke.
The onset of chaos in a fluid flow model of Lorenz. \textit{Ann. N.
Y. Acad. Sci.} \textbf{316}, 400 (1979) [DOI:
10.1111/j.1749-6632.1979.tb29484.x].\vspace*{0.5mm}
\bibitem{55} J.L. Kaplan, J.A. Yorke. \textit{Functional Differential Equations and Approximations of Fixed Points}, edited by H.O.~Peitgen, H.O.~Walther (Springer,  1979), p. 204.\vspace*{0.5mm}
\bibitem{56} A.G. Dorofeev, M.V. Glagolev, T.F. Bondarenko,  N.S.~Pa\-ni\-kov.
Unusual growth kinetics of Arthrobacter globiformis
 and its explanation.  \textit{Mikrobiol.} \textbf{61}, 33 (1992).\vspace*{0.5mm}
\bibitem{57} Ya.B. Pesin.
Characteristic  Lyapunov  indices  and  the  ergodic  theory.
 \textit{Usp. Mat. Nauk} \textbf{32}, No.~4, 55 (1977).\vspace*{5mm}
\begin{flushright}
{\footnotesize Received 00.01.17}
\end{flushright}
\end{thebibliography}
\end{document}